\documentclass[aps,amsmath,amssymb,showpacs,showkeys]{revtex4-2} 
\usepackage[dvips]{graphicx}
\usepackage{times}
\usepackage{multirow}
\usepackage{xcolor}
\usepackage{orcidlink}
\usepackage{hyperref}
\hypersetup{
  colorlinks=true,
  urlcolor=blue,
  linkcolor=red,
  citecolor=blue
}

\begin{document}
\title{Quasinormal modes, thermodynamics and shadow of black holes in 
Hu-Sawicki $\boldsymbol{f(R)}$ gravity theory}

\author{Ronit Karmakar  \orcidlink{0000-0002-9531-7435}}
\email[Email: ]{ronit.karmakar622@gmail.com}

\affiliation{Department of Physics, Dibrugarh University,
Dibrugarh 786004, Assam, India}

\author{Umananda Dev Goswami  \orcidlink{0000-0003-0012-7549}}
\email[Email: ]{umananda@dibru.ac.in}

\affiliation{Department of Physics, Dibrugarh University,
Dibrugarh 786004, Assam, India}

%\date{}
\begin{abstract}
We derive novel black hole solutions in a modified gravity theory, namely the 
Hu-Sawicki model of $f(R)$ gravity. After obtaining the black hole solution, we 
study the horizon radius of the black hole from the metric and then analyse 
the dependence of the model parameters on the horizon. We then use the 6th 
order WKB method to study the quasinormal modes of oscillations (QNMs) of the 
black hole perturbed by a scalar field. The dependence of the amplitude and 
damping part of the QNMs are analysed with respect to variations in model 
parameters and the error associated with the QNMs are also computed. After 
that we study some thermodynamic properties associated with the black hole 
such as its thermodynamic temperature as well as greybody factors. It is 
found that the black hole has the possibility of showcasing negative 
temperatures and is thermodynamically unstable for feasible values of 
model parameters. Then we analyse the geodesics and derive the photon sphere 
radius as well as the shadow radius of the black hole. The photon radius is 
independent of the model parameters while shadow radius showed fair amount of 
dependence on the model parameters. We tried to constrain the parameters with 
the help of Keck and VLTI observational data and obtained some bounds on 
$m$ and $c_{2}$ parameters.
\end{abstract}

%\pacs{04.30.Tv, 04.50.Kd}

\maketitle
\section{Introduction}\label{sec1}
General relativity (GR) has undoubtedly been successful in accounting for 
observational results in the solar system and beyond \cite{1,2}. GR theory 
predicts with great accuracy the precession of the perihelion of planet Mercury 
\cite{1,3} and bending of light due to gravitational field \cite{4} in the 
local as well as distant observations. GR has predicted the existence of 
black holes and gravitational waves (GWs) which has been recently 
experimentally verified by the LIGO-Virgo collaboration \cite{5,6,7,8,9}. The 
recent direct images of the black hole shadows published by the Event Horizon 
Telescope (EHT) group \cite{10,11,12,13,14,15} also back GR in terms of 
experimental verification of the theory. In spite of these successes, GR fails 
to address recent observations like the accelerated expansion of the Universe 
\cite{16,17,18}. Indeed, it does not provide any insights regarding the dark 
components of the Universe \cite{19,20}. Thus, to overcome these issues, 
physicists worked on modified theories of gravity, the most common among them 
includes the $\Lambda$CDM \cite{21} model, $f(R)$ gravity theory 
\cite{22,23,24,25}, $f(R,T)$ gravity theory \cite{26}, $f(Q)$ gravity theory 
\cite{27,28,29,30} and so on (see \cite{31}). These theories can compensate 
for the effects of dark components \cite{32}, explain galactic rotation curves 
\cite{33}, accelerated expansion of the Universe \cite{34,35} and are well 
constrained by modern observations. Rastall gravity proposed in 1972 is 
arguably a unique modified theory of gravity which does not follow from 
an established Lagrangian formalism \cite{36}. It advocates the violation of 
conservation of energy-momentum tensor $T_{\mu\nu}$ and equates it to the 
derivative of the Ricci scalar $R$. In recent times, a number of $f(R)$ 
gravity models have been proposed, some of them include Starobinsky \cite{37}, 
Hu-Sawicki \cite{38}, Sujikawa \cite{40} and other two models mentioned in
Refs.~\cite{39,41} to name a few. These models have been extensively studied 
in the literature regarding various aspects like their cosmological and 
astrophysical implications \cite{33,42,43,44}, dynamical system analysis 
\cite{45,46}, early Universe mysteries \cite{47} and so on. Similarly, $f(Q)$
gravity has also attracted a lot of attention in the recent times and many 
cosmological studies have been carried out in this theory, for instance see 
Refs.~\cite{46,48,49,50} and references therein.    

The first vacuum solution of the Einstein field equations leading to a black
hole was given by Schwarzschild in 1916 \cite{51}. Since then, a number of 
black hole solutions 
have been proposed from time to time in various frameworks of gravity. Black 
holes are often studied with an engulfing field around them. These fields may 
include quintessence fluid \cite{52,53,54,55}, matter in the form of dust, 
radiation \cite{56}, plasma \cite{57}, dark matter halo \cite{58} and so on. 
These surrounding fields have impacts on various thermodynamic properties, 
quasinormal modes (QNMs), shadow radius etc.~of black holes and have been 
extensively studied in the literature. In a recent paper \cite{53}, 
GUP-improved Schwarzschild-type solution, its thermodynamic properties and 
quasinormal modes have been studied. In another work \cite{59}, a 
Schwarzschild-type black hole in Bumblebee gravity has been considered and 
its thermodynamics and shadow have been studied.  

Black hole solutions have been derived in the framework of $f(R)$ gravity in 
many recent papers. In Ref.~\cite{39}, Saffari and Rahvar derived novel black 
hole solutions in the $f(R)$ framework and also proposed a novel $f(R)$ form 
that is feasible in both local and galactic scales. In Ref.~\cite{60}, the 
authors derived black hole solutions in various $f(R)$ models. In a recent work 
\cite{61}, novel black hole solutions were derived in various $f(R)$ models 
and the authors studied topological and thermodynamic properties of the 
solutions obtained. Motivated by these ongoing researches, we derive novel 
black hole solutions in Hu-Sawicki gravity. Here, we intend to study various 
properties relating to the black hole solutions obtained. Our solution is 
unique in the sense that the black hole solution for the Hu-Sawicki model of 
$f(R)$ gravity has not been worked out before to the best of our knowledge 
and thus we are motivated to study its properties including QNMs, 
thermodynamics as well as shadow radius and greybody factors.   

Further, the black hole's shadow also has gained a fair amount of attention, 
credits to the recently released data and images of the black holes at the 
center of M87 galaxy and Sgr A. This has opened up a new window to constrain 
various theories of gravity and parameter values as well. Recently in 
Ref.~\cite{62}, different parameters of modified theories of gravity have 
been constrained using modern shadow radius data. Another work \cite{63} 
constraints regular black hole parameters with shadow data of the EHT. Recent 
works regarding black hole shadows have gained momentum as shadow provides 
interesting new insights and data to constrain black hole physics \cite{64,
65,66,67,68,69,71,71-1,71-2}.
 
QNMs of oscillations of a perturbed black hole hold promise of 
constraining physics at the extreme regimes of black holes. The QNMs are 
basically complex frequencies linked to GWs produced when a black hole is 
perturbed by some external means. There has been an upsurge of research 
regarding various aspects of QNMs, new techniques of computing QNMs, their 
relationships with shadow data and so on. The WKB method of computing QNMs 
is the most widely used technique, though many spectral and analytical 
techniques are often used in combination. There has been a wide range of 
applications of QNMs in understanding various phenomena such as testing the 
No-Hair theorem \cite{71-3} and constraining theories of modified gravity 
\cite{71-4}. QNMs can also be used to study the stability of background 
spacetime when it is acted upon by a minute perturbation \cite{71-5}. The 
relation between shadow radius and QNMs has been dealt with in 
Ref.~\cite{71-2}. QNMs and Hawking radiation sparsity for GUP-corrected black 
holes with topological defects have been studied in Ref.~\cite{72}. A brief 
account of various methods employed in recent times to compute the QNMs can be 
found in the Refs.~\cite{72,73,74,75,76}.

Black hole thermodynamics has gained momentum and attracted a lot of attention 
following the path-breaking work of Bekenstein and Hawking 
\cite{76-1,76-2,76-3}. Their idea led to the development of four laws of 
black hole thermodynamics. Recently a number of research works have been 
carried out in this field. Schwarzschild black holes with quantum corrections 
have been investigated for scattering and absorption cross-section \cite{76-4}.
In Ref.~\cite{76-5}, the authors studied absorption and scattering 
by a black hole with a global monopole in $f(R)$ gravity. Recently, 
thermodynamic properties of extended GUP-corrected black holes has been 
carried out in Ref.~\cite{76-6}. Thermodynamics of static dilaton black holes 
have been studied in Ref.~\cite{76-7}.     

In this work, we derive black hole solutions in the Hu-Sawicki model of $f(R)$ 
gravity and study its thermodynamic properties along with its QNMs using 
the 6th-order WKB method. We compute the shadow radius and present the plots 
of its variation with respect to different model parameters. The primary 
motivation for choosing the Hu-Sawicki model is that black hole solutions have 
not been worked out in this model, and thus it is really intriguing to study 
the properties of such a solution. The Hu-Sawicki model is a viable choice as 
it is observationally consistent in cosmological scales \cite{76-8,76-9}. It is 
consistent with the solar system tests and thus shows viability in the local 
scales as well \cite{76-10}. Another reason for choosing this model is 
that the model parameters of Hu-Sawicki gravity have not been constrained 
using the available shadow radius data of black holes, although it should be 
noted that data from cosmological observations have been utilised to constrain 
this model's parameters. The choice of this model is thus motivated by the gap 
in the literature and the viability of the model. Some recent articles that 
utilise the Hu-Sawicki model to study various aspects of astrophysics and 
cosmology can be found in Refs.~\cite{38, 76-11,76-12}.
 
The plan of the paper is as follows. 
In the second section, we introduce the field equations in the $f(R)$ gravity 
framework and briefly discuss the method of solving the equations. After 
attaining the black hole solution, we move to the third section where we 
compute QNMs of the black hole. Then in the fourth section, we discuss the 
thermodynamic properties including temperature, entropy and heat capacity 
along with greybody factors. Then in the fifth section, we compute the shadow 
radius and plot it for variations in parameters. Finally, we conclude the work 
with a brief summary and future scopes.    

\section{Field equations in $\boldsymbol{f(R)}$ gravity theory}\label{sec2}
The field equations for the $f(R)$ gravity theory will be presented here in 
the spherically symmetric spacetime by adopting the metric formalism of the
theory, in which the variation of action is done with respect to the metric 
only. The $f(R)$ gravity field equation can be obtained from an action in
which the Ricci scalar $R$ in the Einstein-Hilbert action is replaced by some
function $f(R)$ of $R$. Thus the generic action of the $f(R)$ gravity theory 
can be written as \cite{60}:
\begin{equation}
S=\frac{1}{2\kappa}\int d^4 x\, \sqrt{-g} f(R) + S_m,
\label{eqa1}
\end{equation}
where $\kappa = 8\pi G c^{-4}$ and $S_m$ is the matter part of the action. As 
mentioned already, taking the variation of the above action \eqref{eqa1} with 
respect to the metric $g_{\mu\nu}$, one can obtain the field equations of 
$f(R)$ gravity as 
\begin{equation}
F R_{\mu\nu}-\frac{1}{2}f(R) g_{\mu\nu}-\left(\nabla_{\mu}\nabla_{\nu}-g_{\mu\nu}\square\right)F=\kappa\, T_{\mu\nu},
\label{eqa2}
\end{equation}
where $F=df(R)/dR$ and $\square=\nabla_{\alpha}\nabla^{\alpha}$. Taking the 
trace of this \eqref{eqa2}, we can write the function $f(R)$ as 
\begin{equation}
f(R) = \frac{1}{2}\left(3\,\square F + FR - \kappa T\right).
\label{eqa4}
\end{equation}
The derivative of this Eq.~\eqref{eqa4} with respect to the 
radial coordinate $r$ leads to an equation in terms of $F$ and $R$ as 
given by
\begin{equation}
F' R-F R'+3(\square F)'=\kappa T', 
\label{eqa5}
\end{equation}
where the prime denotes the derivative with respect to the radial coordinate
$r$. This equation will serve as a consistency relation for the function $F$ 
that any solution for $F$ must satisfy this relation in order to be a solution
of the field equations, Eq.~\eqref{eqa2}. Further, using Eq.~\eqref{eqa4} 
in Eq.~\eqref{eqa2}, the field equations can be expressed in terms $F$ instead
of $f(R)$ as
\begin{equation}
R_{\mu\nu}-\frac{1}{4}\,g_{\mu\nu}R=\frac{\kappa}{F}\Big(T_{\mu\nu}-\frac{1}{4}\,g_{\mu\nu}T\Big)+\frac{1}{F}\Big(\nabla_{\mu}\nabla_{\nu} F-\frac{1}{4}\,g_{\mu\nu}\square F\Big).
\label{eqa6}
\end{equation}
Considering the case of the vacuum where the energy-momentum tensor and its 
trace vanish, we can rewrite the above equation as 
\begin{equation}
F R_{\mu\nu}-\nabla_{\mu}\nabla_{\nu} F=\frac{1}{4}\,g_{\mu\nu}\Big(F R-\square F\Big).
\label{eqa7}
\end{equation}

\begin{figure}[h!]
\includegraphics[scale=0.7]{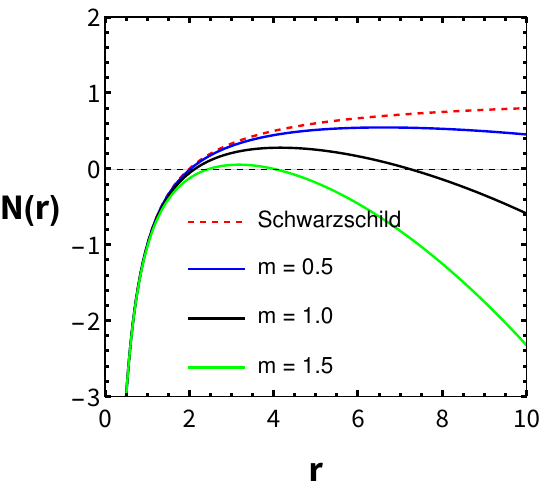} \hspace{1cm}
\includegraphics[scale=0.7]{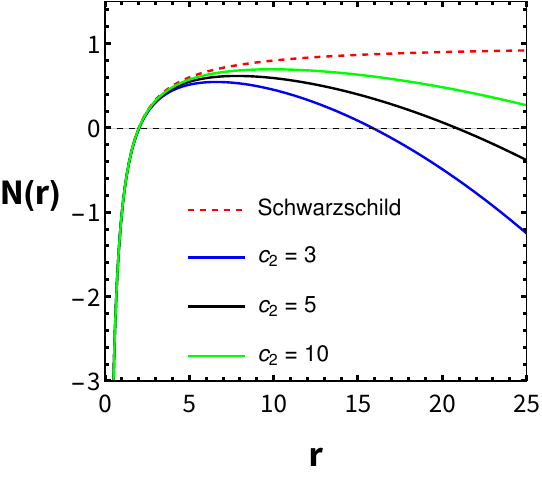}
\vspace{-0.2cm}
\caption{Black hole metric function versus radial distance $r$ for different 
values of parameters. In the left plot, we use $c_2 = 3$ and $n = 1$ while in 
the right plot, we use $m = 0.5$ and $n = 1$. The red dashed line shows the 
ideal Schwarzschild case.}
\label{fig01}
\end{figure}

Since we are interested in the solution of this time-independent spherically 
symmetric vacuum field equations following the procedure adopted in 
Ref.~\cite{60}, we consider a generic spherically symmetric metric in 
the form:
\begin{equation}
g_{\mu\nu}=\begin{pmatrix}
-N(r) & 0 & 0 & 0 \\ 
0 &  M(r) & 0 & 0 \\ 
0 & 0 &  r^2 & 0\\ 
0 & 0 & 0 & r^2 \sin^2 \theta \\
\end{pmatrix},
\label{eqa3}
\end{equation}
where $N(r)$ and $M(r)$ are metric coefficients to be determined, associated 
with the time and space components of the metric respectively which are indeed 
functions of $r$. For this spherically symmetric metric, both sides of 
Eq.~\eqref{eqa7} become diagonal and accordingly, we can define an index 
independent parameter from this equation as 
\begin{equation}
P_{\mu}\equiv \frac{F R_{\mu\mu}-\nabla_{\mu}\nabla_{\mu} F}{g_{\mu\mu}}.
\label{eqa8}
\end{equation}
As this quantity $P_{\mu}$ is independent of indices, we can have 
$P_{\mu}-P_{\nu}=0$ for all $\mu$ and $\nu$ values and hence from this 
property one can obtain the following expressions:
\begin{align}
2F \frac{X'}{X}+r F' \frac{X'}{X}-2\, r F'' & = 0,
\label{eqa9}\\[8pt]
%\end{equation}
%\begin{equation}
N''+\left(\frac{F'}{F}-\frac{X'}{2X}\right)N'-\frac{2}{r}\left(\frac{F'}{F}-\frac{X'}{2X}\right)N-\frac{2}{r^2}N + \frac{2}{r^2}& = 0.
\label{eqa10}
\end{align}
Here $X=MN$. In this work, our solution is considered to have constant 
curvature for the sake of simplicity. Hence the terms $F'$ and $F''$ 
vanish and the field Eqs.~\eqref{eqa9} and \eqref{eqa10} take the forms:
\begin{align}
N M'+N' M & = 0,
\label{eqa11}\\[8pt]
%\end{equation}
%\begin{equation}
1-M+\frac{r}{2}\left(\frac{N'}{N}+\frac{M'}{M}\right)\left(\frac{r}{2} \frac{N'}{N}-1\right)-\frac{r^2 N''}{2N} & = 0.
\label{eqa12}
\end{align}
Solving these two Eqs.~\eqref{eqa11} and \eqref{eqa12}, one can obtain:
\begin{equation}
M(r)=\frac{s_1}{N(r)},\;\;\; \text{and}\;\;\; 
N(r)=s_1 + \frac{s_2}{r}+s_3\, r^2,
\label{eqa13}
\end{equation}
where $s_1$, $s_2$ and $s_3$ are constants of integration. In order to get 
these coefficients, we follow the procedure in Refs.~\cite{60,61} and 
compare the second solution with the standard Schwarzschild-de Sitter 
solution. The standard Schwarzschild-de Sitter black hole metric coefficient 
is \cite{60} 
\begin{equation}
C(r)=1-\frac{2M}{r}-\frac{\Lambda r^2}{3}.
\label{eqa15}
\end{equation}
Again, the relationship between the scalar curvature and cosmological 
constant is \cite{60}
\begin{equation}
R=-\,4\Lambda.
\label{eqa16}
\end{equation}
Now, comparing the second solution in Eq.~\eqref{eqa13} with 
Eq.~\eqref{eqa15}, we have
\begin{equation*}
s_1 = 1, s_2 = -\,2M, s_3 = -\,\frac{\Lambda}{3} = \frac{R}{12}.
\end{equation*}
From \eqref{eqa4}, considering the vacuum case and constant curvature $R_0$, 
we have
\begin{equation}
R_0=\frac{2f(R_0)}{F(R_0)}.
\label{eqa17}
\end{equation}
As mentioned earlier, the $f(R)$ gravity model we employed in our work is the 
Hu-Sawicki model \cite{38}, which is given by
\begin{equation}
f(R)=-\,m^2\frac{c_1(\frac{R}{m^2})^n}{c_2 (\frac{R}{m^2})^n +1},
\label{eqa18}
\end{equation}
where $m$, $n\, (>0)$, $c_1$ and $c_2$ are the model parameters. Here $c_1$ and 
$c_2$ are dimensionless and $m$ represents the mass (energy) scale 
\cite{32}. For this model, we solve Eq.~\eqref{eqa17} to get the
constant curvature
\begin{equation}
R_0=12 s_3=m^2 \bigg(\frac{n-2}{2\,c_2}\bigg)^\frac{1}{n}\!\!\!.
\label{eqa19}
\end{equation}
Thus, we arrive at our black hole solution for the Hu-Sawicki model as
\begin{equation}
N(r)=1-\frac{2M}{r}+ \frac{m^2}{12} \bigg(\frac{n-2}{2c_2}\bigg)^\frac{1}{n}r^2.
\label{eqa20}
\end{equation}
It is clear that our black hole solution is independent of the Hu-Sawicki 
model parameter $c_1$. Fig~\ref{fig01} shows the metric function versus 
radial distance for the other two Hu-Sawicki model parameters $m$ and $c_2$, 
while taking the parameter $n=1$ for simplicity (this is considered for the 
whole study if we do not mention otherwise). In the plots, it is seen that 
the black hole solution \eqref{eqa20} has two horizons for a range of parameter 
values. The first plot shows that with increasing $m$, the outer horizon moves 
closer to the inner horizon, while the second plot shows that for higher 
values of $c_2$, the outer horizon increases. After a certain higher value of 
the parameter $m$ and the lower value of the parameter $c_2$, the black hole 
appears to be a horizonless singularity for the given values of the other 
parameters.    

\section{Quasinormal modes of the Black Hole}\label{sec3}
In this section, we compute the QNMs of the black hole \eqref{eqa20} using the 
most common method, the 6th-order WKB approximation method. To this end, we 
apply a perturbation to the black hole in the form of a probe coupled minimally
to a scalar field $\Phi$ and having the equation of motion \cite{54}:
\begin{equation}
\frac{1}{\sqrt{-g}}\,\partial_{\alpha}\Big(\sqrt{-g}g^{\alpha\beta}\partial_{\beta}\Big)\Phi=\mu^2 \Phi,
\label{eqb1}
\end{equation}
where $\mu$ is the mass of the scalar field, which for our convenience will be 
taken as a massless scalar field with $\mu=0$. We can express the scalar field 
$\Phi$ in terms of spherical harmonics of the form \cite{54}:
\begin{equation}
\Phi(t,r,\theta,\phi)=e^{-i \omega t}\frac{\Psi(r)}{r}\,Y_l^p(\theta,\phi).
\label{eqb2}
\end{equation}
Here $\Psi(r)$ represents the radial part of the wave and $Y_l^p$ represents 
the spherical harmonic part. Employing Eq.~\eqref{eqb2} in Eq.~\eqref{eqb1}, 
we get a Schr\"odinger-type equation, as given below:
\begin{equation}
\frac{d^2\Psi}{dx^2}+\left[\omega^2-V(x)\right]\Psi=0,
\label{eqb3}
\end{equation}
with the new variable, the tortoise coordinate, which is defined as
\begin{equation}
x=\int \frac{dr}{N(r)}.
\label{eqb4}
\end{equation}
The effective potential in Eq.~\eqref{eqb3} can be expressed as 
\begin{equation}
V(r)=N(r)\bigg(\frac{N'(r)}{r}+\frac{l(l+1)}{r^2}\bigg).
\label{eqb6}
\end{equation} 
It is necessary to apply the appropriate boundary conditions to 
Eq.~\eqref{eqb3} for the physical consistency both at the black hole horizon
and at infinity. For spacetime which is flat asymptotically the following 
quasinormal criteria have to be satisfied:
\begin{equation}
\Psi(x) \rightarrow \begin{cases} A e^{+i \omega x} \;\;\; \text{if} \;\; x \rightarrow -\infty,\\
B e^{-i \omega x} \;\;\; \text{if} \;\; x \rightarrow +\infty \end{cases}.
\label{eqb5}
\end{equation}
Here, the coefficients $A$ and $B$ represent the amplitudes of the waves. 
These ingoing and outgoing waves are in accordance with the physical 
requirements that nothing can escape from the black hole horizon and no 
radiation comes from the infinity respectively. Further, these make sure of 
the existence of an infinite set of discrete complex numbers, usually known as 
the QNMs.   

\begin{figure}[h!]
\includegraphics[scale=0.6]{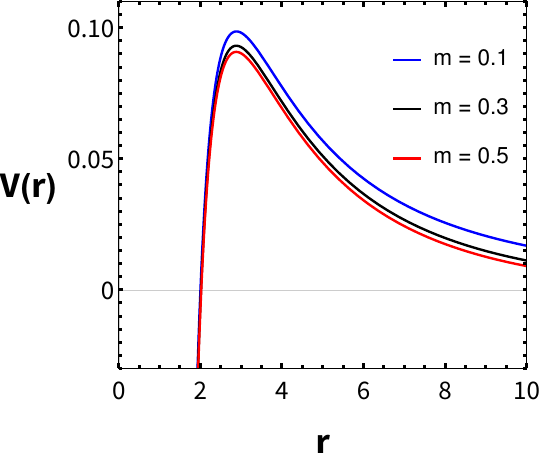}\hspace{3mm}
\includegraphics[scale=0.6]{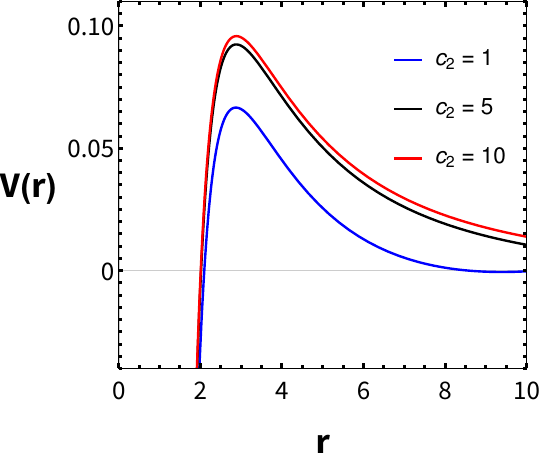}\hspace{3mm}
\includegraphics[scale=0.6]{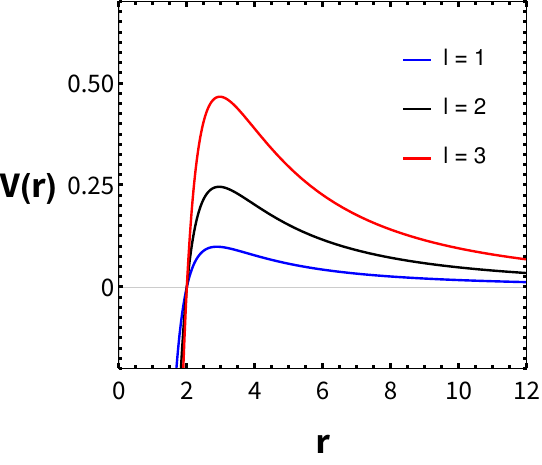}
\vspace{-0.2cm}
\caption{Behaviours of black hole potential with respect to radial distance 
$r$ for different model parameters. The left plot uses values of $c_{2} = 2$, 
$n = 1$ and multipole $l = 1$. The middle plot uses $m = 0.5$, $n = 1$ and 
multipole $l = 1$. The right plot uses $m = 0.1$, $n = 1$ and $c_{2}= 2$.}
\label{fig02}
\end{figure}

To study the behaviour of the potential \eqref{eqb6} before calculating the
QNMs of the black hole \eqref{eqa20}, we plot the potential versus $r$ for 
different variations of model parameters in Fig.~\ref{fig02}. As seen from the 
left plot of Fig.~\ref{fig02}, the peak of the potential decreases for higher 
$m$ values. From the middle plot, one can see that increasing values of 
parameter $c_2$ enhances the peak of the potential. A similar trend is seen 
with mutipole $l$ values, where peaks are found to increase for higher 
$l$ values.

\begin{figure}[h!]
\includegraphics[scale=0.35]{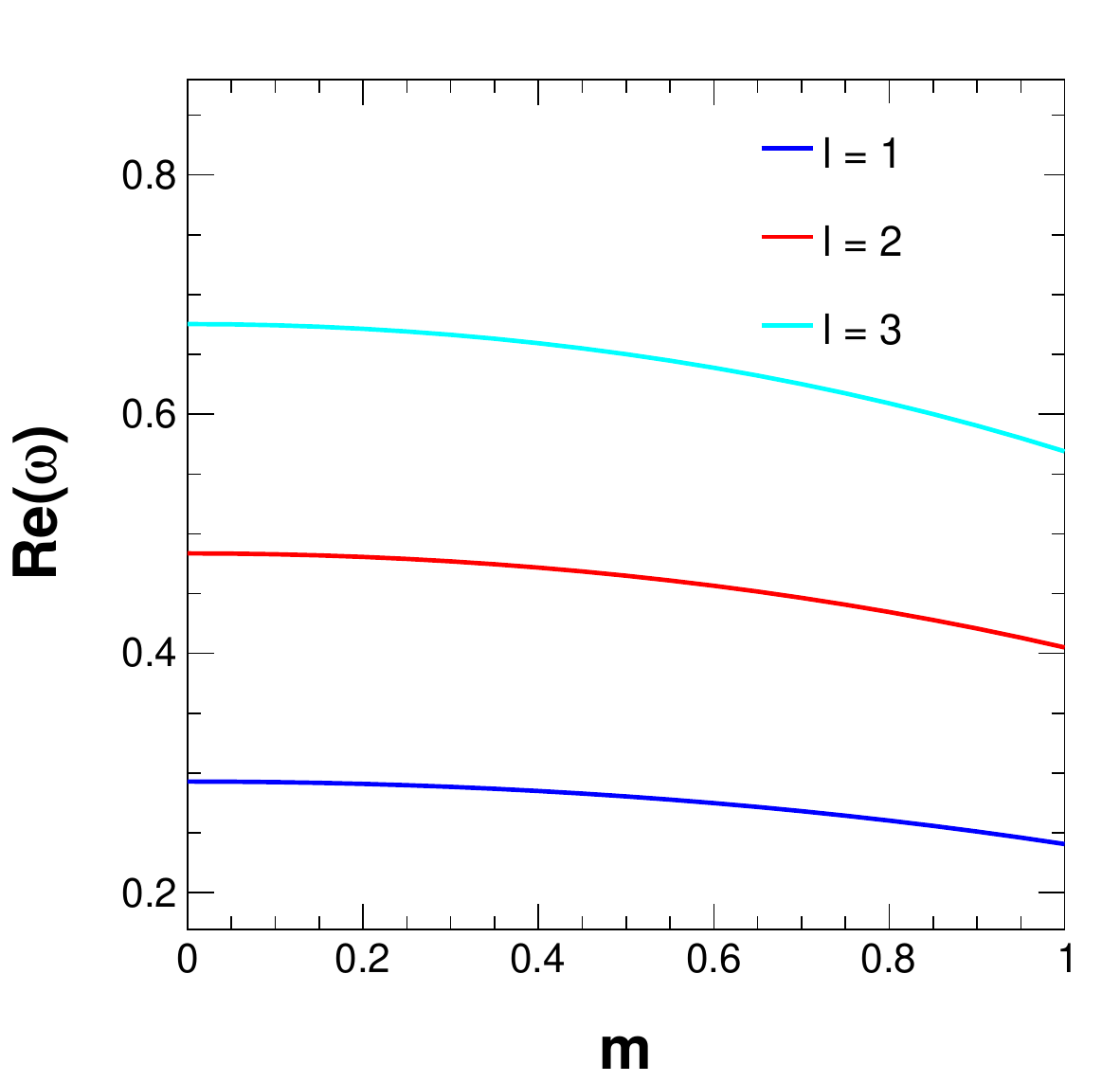}\hspace{1cm}
\includegraphics[scale=0.35]{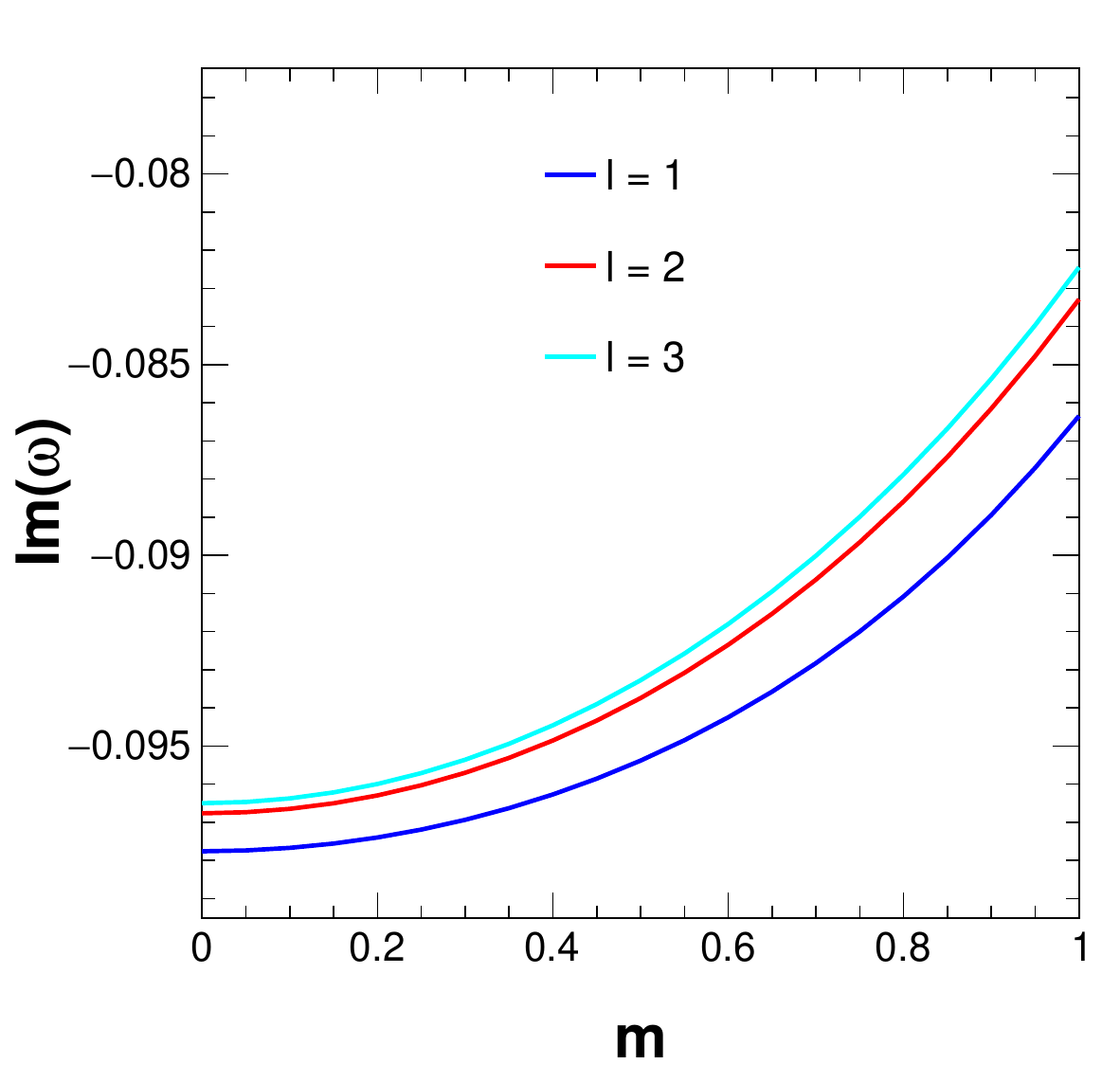}
\vspace{-0.2cm}
\caption{Variation of amplitude and damping of QNMs with respect to parameter 
$m$ for three values of multipole $l$. Here we use $n=1$, $M=1$ and $c_2=4$ to 
represent the features of QNMs.}
\label{fig03}
\end{figure}

\begin{figure}[h!]
\includegraphics[scale=0.35]{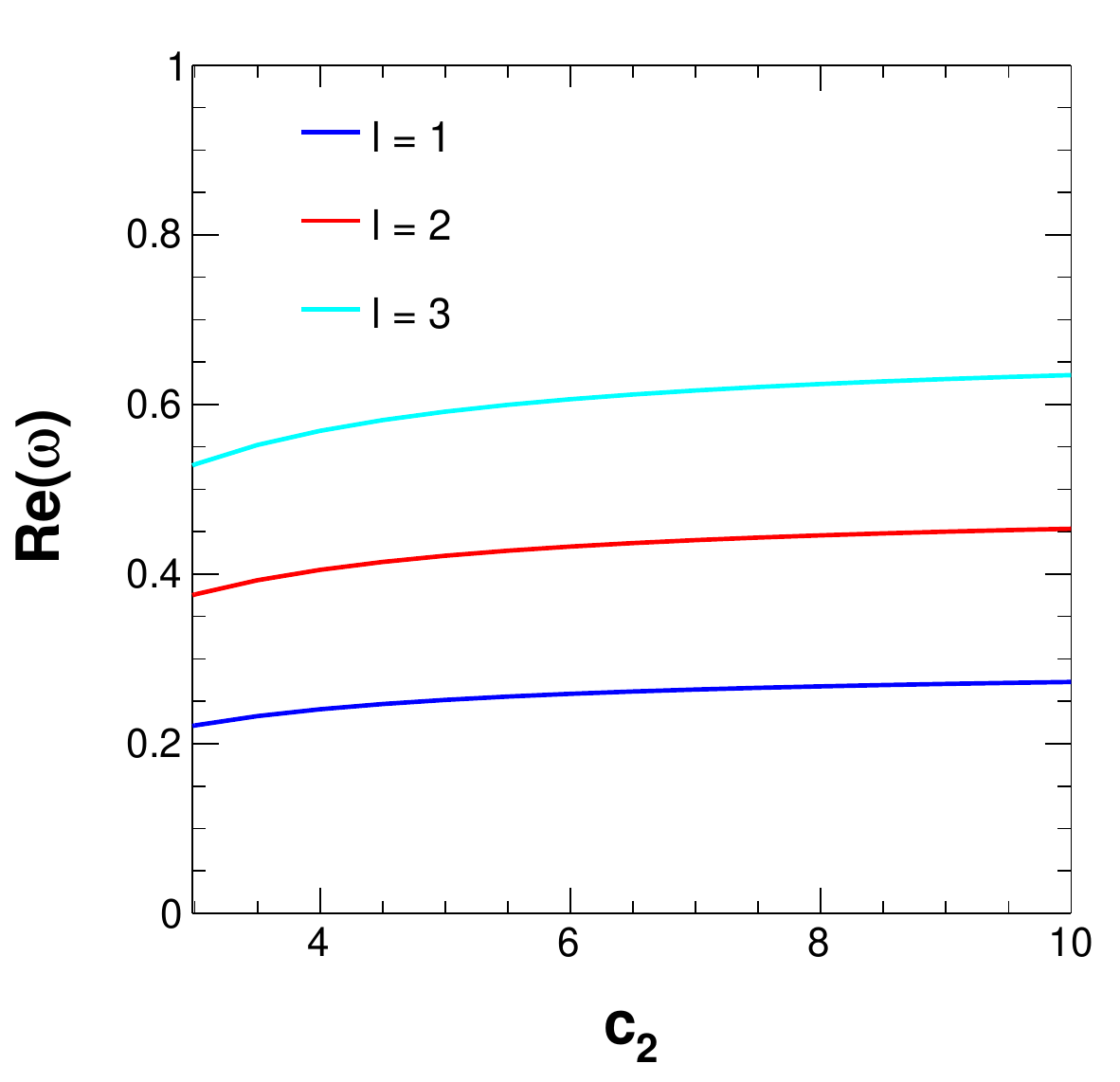}\hspace{1cm}
\includegraphics[scale=0.35]{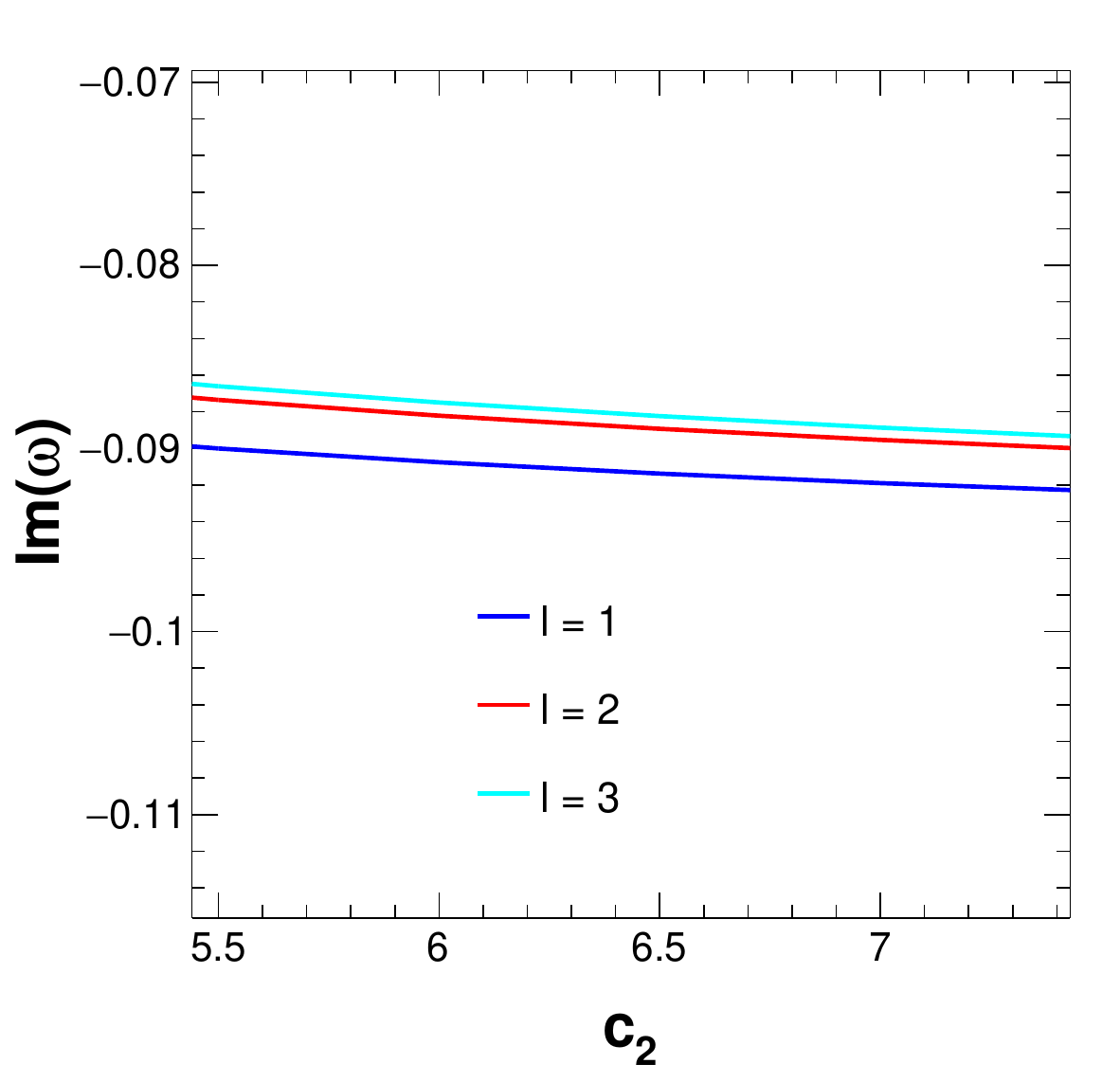}
\caption{Variation of amplitude and damping of QNMs with respect to parameter 
$c_2$ for three values of multipole $l$. Here parameters $n=1$, $M=1$ and 
$m=1$ have been used.}
\label{fig04}
\end{figure}

The QNMs have been calculated utilising the 6th-order WKB method in the form 
of their amplitude and damping varying with the model parameters. As seen from 
Fig.~\ref{fig03}, the general trend of amplitude and damping of QNMs is that 
both decrease with the parameter $m$ for all values of multiple $l$. On the 
other hand, Fig.~\ref{fig04} shows that both amplitude and damping increase 
slightly with the parameter $c_2$ for all $l$ values. In both cases, the 
effect of $l$ is more dominating on the amplitude than that on the damping. 
Moreover, in both cases of $m$ and $c_2$ variations, the amplitude increases, 
while the damping decreases with the increasing value of $l$.   

We compute the error associated with the WKB QNMs with a prescribed formula 
that has been used extensively in the literature. This error estimating formula 
for the WKB method is as follows \cite{54,72,73}:
\begin{equation}
\Delta_6=\frac{\big|W\!K\!B_7-W\!K\!B_5\big|}{2},
\label{eqb7}
\end{equation} 
where $W\!K\!B_5$ and $W\!K\!B_7$ are respectively the QNMs obtained from the 
5th and 7th order WKB method. In Table \ref{tab01}, we present the 6th-order 
WKB QNMs along with the associated errors for various values of the model 
parameters along with the multipole number $l$. It is clear that the errors 
are reduced for higher multipole numbers $l$. Similar trends in variations of
QNMs with respect to different parameters as seen in Figs.~\ref{fig03} and
\ref{fig04} are displayed in the tabulated data. The estimated errors in most 
of the cases lie around $10^{-4}-10^{-5}$.

\begin{table}[h!]
\caption{6th-order WKB QNMs of the black hole specified by the metric 
function \eqref{eqa20} for the multipoles $l=1,2,3$ with $n=0$ and for 
different values of the model parameters. The estimated errors associated with 
the WKB results have also been shown. The QNMs with $m=c_2=0$ 
represent the Schwarzschild case, which are listed for the comparision 
purpose only.}
\vspace{8pt}
\centering
\begin{tabular}{c@{\hskip 5pt}c@{\hskip 10pt}c@{\hskip 10pt}c@{\hskip 10pt}c@{\hskip 10pt}c@{\hskip 10pt}c}
\hline \hline
\vspace{2mm}
& Multipole & $m$ & $c_2$ & 6th order QNMs & $\Delta_{6}$  \\
\hline
&\multirow{4}{4em}{$l=1$} &  $0.0$ & $0.0$ & 0.292910 - 0.097762i & $0.9824 \times 10^{-6}$ &\\
& & $0.1$ & $4.0$ & 0.292420 - 0.097672i  & $0.9871 \times 10^{-4}$ &\\ 
& & $0.3$ & $4.0$ & 0.288479 - 0.096937i & $0.9483 \times 10^{-4}$ &\\
& & $0.5$ & $4.0$& 0.280478 - 0.0953844i  & $0.9076 \times 10^{-4}$ &\\ 
& & $1.0$ & $4.0$ & 0.240646 - 0.0863446i & $0.6341 \times 10^{-4}$ &\\
& & $1.0$ & $6.5$ & 0.261572 - 0.0913734i & $0.7809 \times 10^{-4}$ &\\
 \hline \hline
&\multirow{4}{4em}{$l=2$} & $0.0$ & $0.0$ & 0.483642 - 0.096766i & $0.8364 \times 10^{-7}$ &\\
& & $0.1$ & $4.0$ & 0.482906 - 0.0966493i  & $0.8355 \times 10^{-5}$ &\\ 
& & $0.3$ & $4.0$ & 0.476986 - 0.0957027i  & $0.7845 \times 10^{-5}$ &\\ 
& & $0.5$ & $4.0$ & 0.46497 - 0.0937446i  & $0.7326 \times 10^{-5}$ &\\ 
& & $1.0$ & $4.0$ & 0.405024 - 0.0832849i & $0.5689 \times 10^{-5}$ &\\
& & $1.0$ & $6.5$ & 0.436572 - 0.0889274i & $0.6718 \times 10^{-5}$ &\\
 \hline \hline
&\multirow{4}{4em}{$l=3$} & $0.0$ & $0.0$ & 0.675366 - 0.096501i & $0.1032 \times 10^{-7}$ &\\
& & $0.1$ & $4.0$ & 0.674374 - 0.096375i  & $0.1534 \times 10^{-5}$ &\\ 
& & $0.3$ & $4.0$ & 0.666393 - 0.095361i  & $0.1534 \times 10^{-5}$ &\\ 
& & $0.5$ & $4.0$ & 0.65018 - 0.0932773i  & $0.1527 \times 10^{-5}$ &\\ 
& & $1.0$ & $4.0$ & 0.568930 - 0.082446i & $0.5235 \times 10^{-6}$ &\\
& & $1.0$ & $6.5$ & 0.611771 - 0.088233i & $0.1033 \times 10^{-5}$ &\\
\hline \hline \vspace{4mm}
\end{tabular}
\label{tab01}
\end{table}
    
\section{Thermodynamic Charactistics of the black hole}\label{sec4}
\begin{figure}[h!]
\includegraphics[height=6cm,width=6cm]{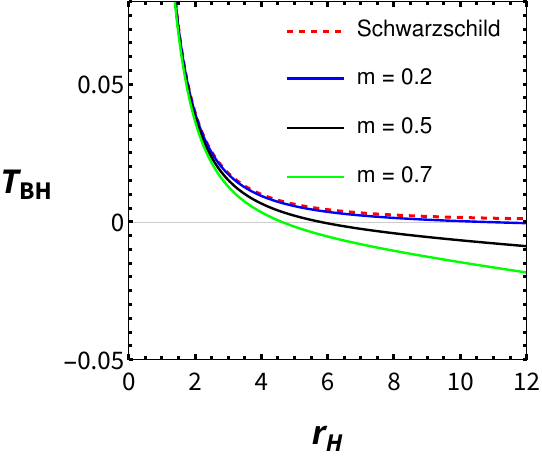}\hspace{1cm}
\includegraphics[height=6cm,width=6cm]{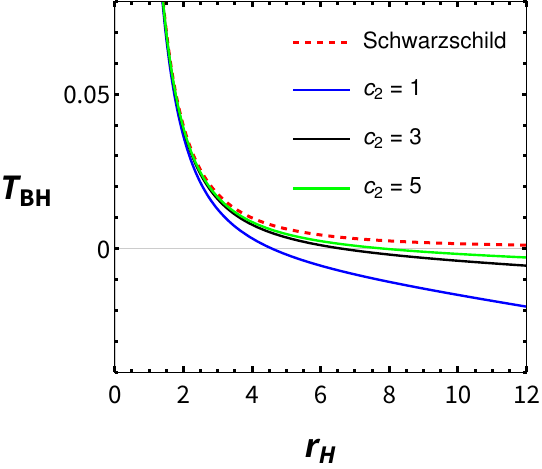}
\vspace{-0.2cm}
\caption{Variation of temperature versus $r_H$ for three different values of 
$m$ on the left plot and for three different values of $c_{2}$ on the right 
plot. For the left plot, we use $n = M = 1$ and $c_{2} = 2$, while for the 
right plot, we use $M = n = 1$ and $m = 0.5$.}
\label{fig05}
\end{figure}

As mentioned earlier, a black hole as a thermodynamic system was first 
conceptualised in the ground-breaking work of Hawking and Bekenstein in the 
early 1970s. In this paper, we analyse the black hole temperature and the grey 
body factors which are important properties that give useful insights in this 
regard. The temperature of a black hole is an important property that is 
associated with the quantum particles created near its horizon. It is 
inversely related to the size or mass of the black hole, that is a larger 
black hole will have a lower temperature. Hawking conceptualised the 
temperature of black hole in the form of radiation, which is today referred to 
as Hawking radiation. It remains a challenge to detect such radiation 
experimentally. We can theoretically compute the black hole temperature from 
the metric solution \eqref{eqa20}, by employing the simple relation:
\begin{equation}
T_{BH}=\frac{N'(r)}{4\pi} =\frac{1}{4\pi r_H^2}\bigg[2M+\frac{m^2}{12}\Big(\frac{n-2}{c_2}\Big)^{\frac{1}{n}}r_H^3\bigg].
\label{eqb8}
\end{equation}
It can also be calculated using the First law of black hole thermodynamics 
as follows:
\begin{equation}
T_{BH}=\frac{dM}{dS}=\frac{dM'}{dS'} = \frac{1}{4\pi r_H^2}\bigg[2M+\frac{m^2}{12}\Big(\frac{n-2}{c_2}\Big)^{\frac{1}{n}}r_H^3\bigg].
\label{eqb9}
\end{equation}
This confirms our computation of the black hole temperature from the first 
law. We plot the thermodynamic temperature \eqref{eqb8} with respect to the 
horizon radius $r_H$ in Fig.~\ref{fig05}. Here, we see clearly that with 
horizon radius $r_H$, the temperature of the black hole is always in the 
decreasing trend. The left plot shows the temperature variations with respect 
to $r_H$ for three different values of parameter $m$. It is seen that higher 
values of $m$ lead to negative temperatures. While for the parameter 
$c_2$, lower values lead to negative temperatures as can be seen from the 
right panel of Fig.~\ref{fig05}. Though the negative temperature seems 
unphysical, this has been encountered in the literature and explained as a 
possible state of formation of ultra-cold black holes \cite{54,76}.  Negative 
temperatures may also suggest that the black hole is thermodynamically unstable \cite{nt}.

\begin{figure}
\includegraphics[scale=0.7]{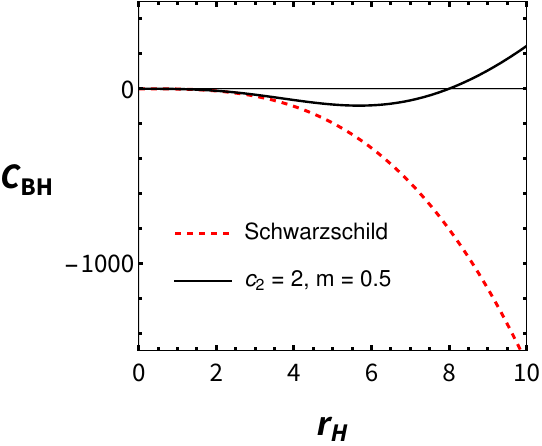}\hspace{0.6cm}
\includegraphics[scale=0.7]{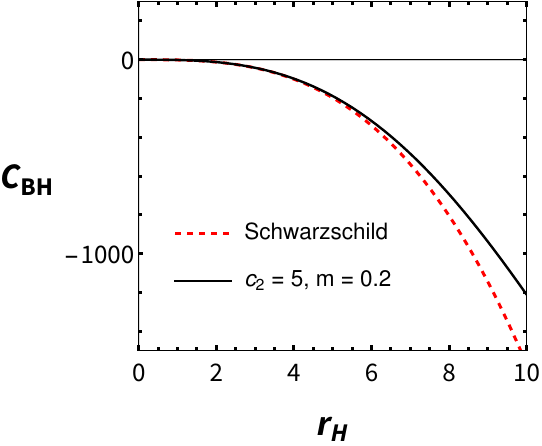}
\caption{Heat capacity of the black hole versus horizon radius is 
shown. The left plot shows that for $r_H\ge8$, the heat capacity takes 
positive values, otherwise the heat capacity remains negative. It indicates 
the black hole is thermodynamically unstable in most of the parameter space.}
\label{fig051}
\end{figure}

To study the thermodynamic stability of the black hole solution, we 
have computed the heat capacity of the solution. Following the first law of 
black hole thermodynamics, the formula for calculating the heat capacity of the 
black hole can be found as:
\begin{equation}
C_{BH}=\frac{dM}{dT}=3 \pi  r^2 \left(\frac{4 (6 M+r)}{m^2 r^3 \left(\frac{\frac{n}{2}-1}{\text{c2}}\right)^{1/n}-24 M}+1\right).
\label{eqb91}
\end{equation}
The plots of the heat capacity of the black hole for different values 
of the model's parameters are shown in Fig.~\ref{fig051}. Here we have 
considered the constraint values of the model's parameters that are obtained 
from the shadow analysis as can be seen from Fig.~\ref{fig08}. In the left 
hand plot of Fig.~\ref{fig051}, we have used $m=0.5$ and $c_2 =2$, which are 
the extreme bounds obtained from shadow radius. It is seen that when 
$r_H\geq 8$, the heat 
capacity takes positive values, otherwise it remains negative. In the right 
hand plot of the same figure, we have used intermediate values of the 
parameters in the viable range suggested by shadow analysis. It is seen from 
this plot that the heat capacity always remains negative, which indicates 
thermodynamic instability. Thus we may comment that the black hole is mostly 
thermodynamically unstable.

\begin{figure}[h!]
\includegraphics[scale=0.27]{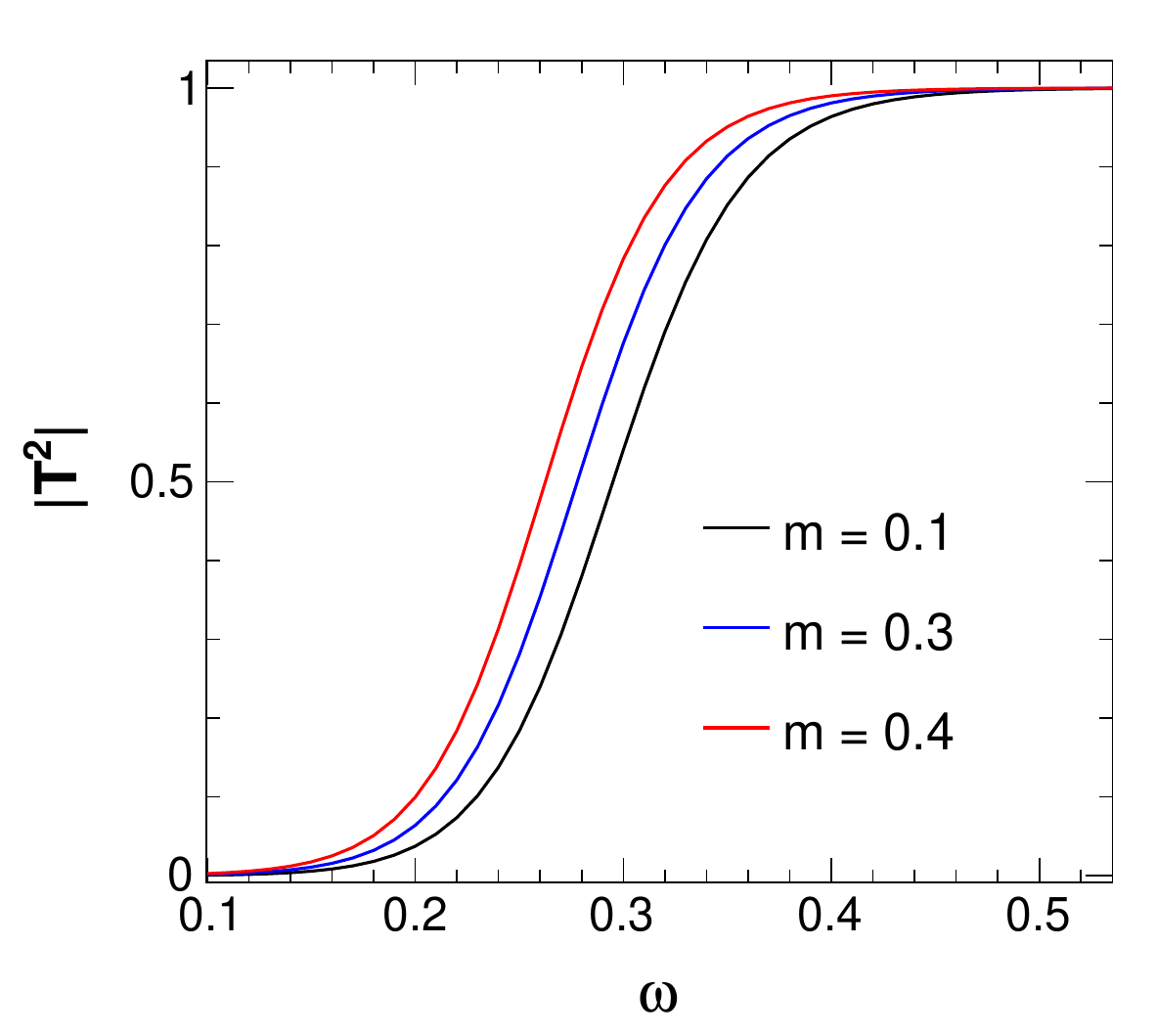}\hspace{0.3cm}
\includegraphics[scale=0.27]{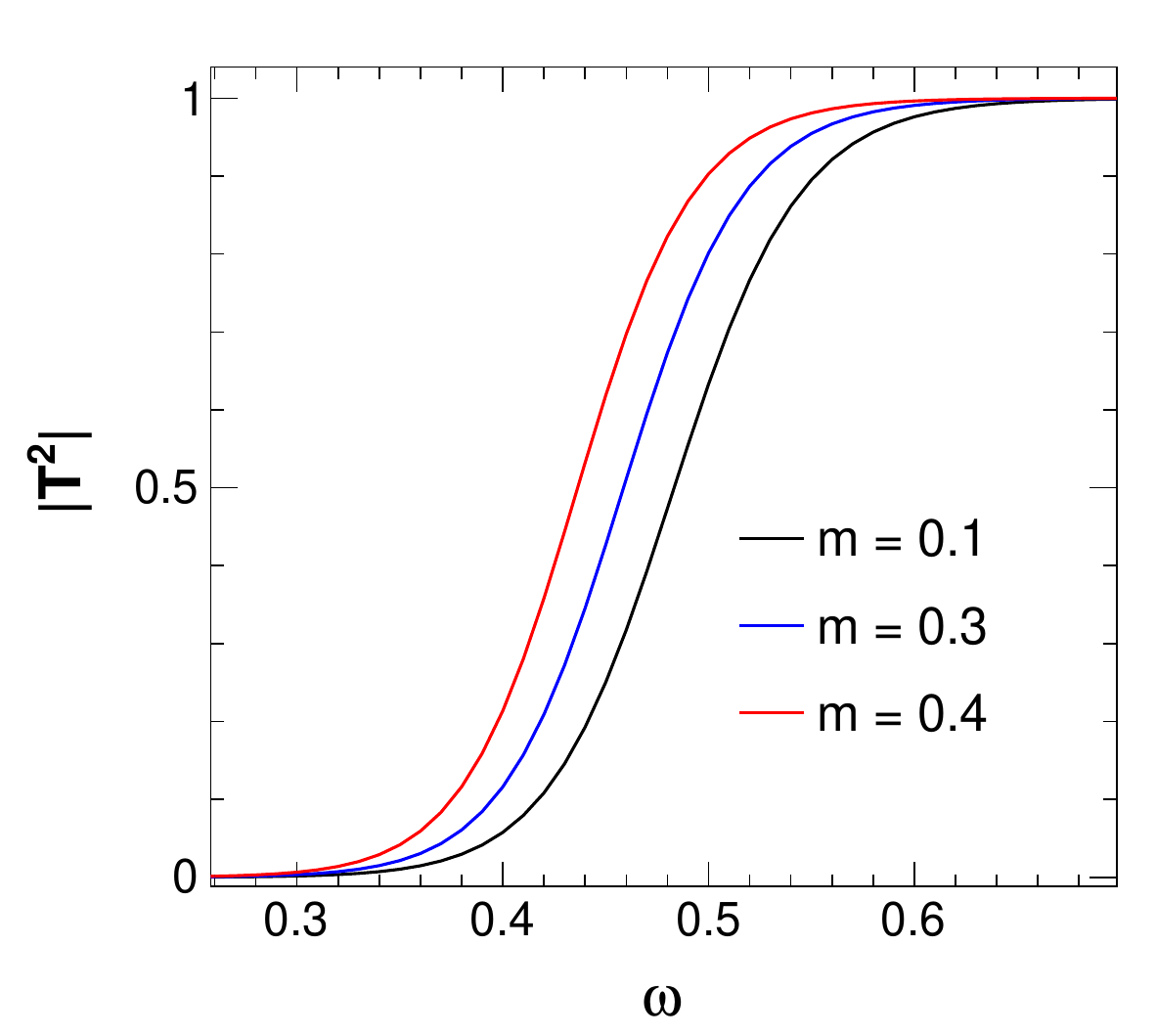}\hspace{0.3cm}
\includegraphics[scale=0.272]{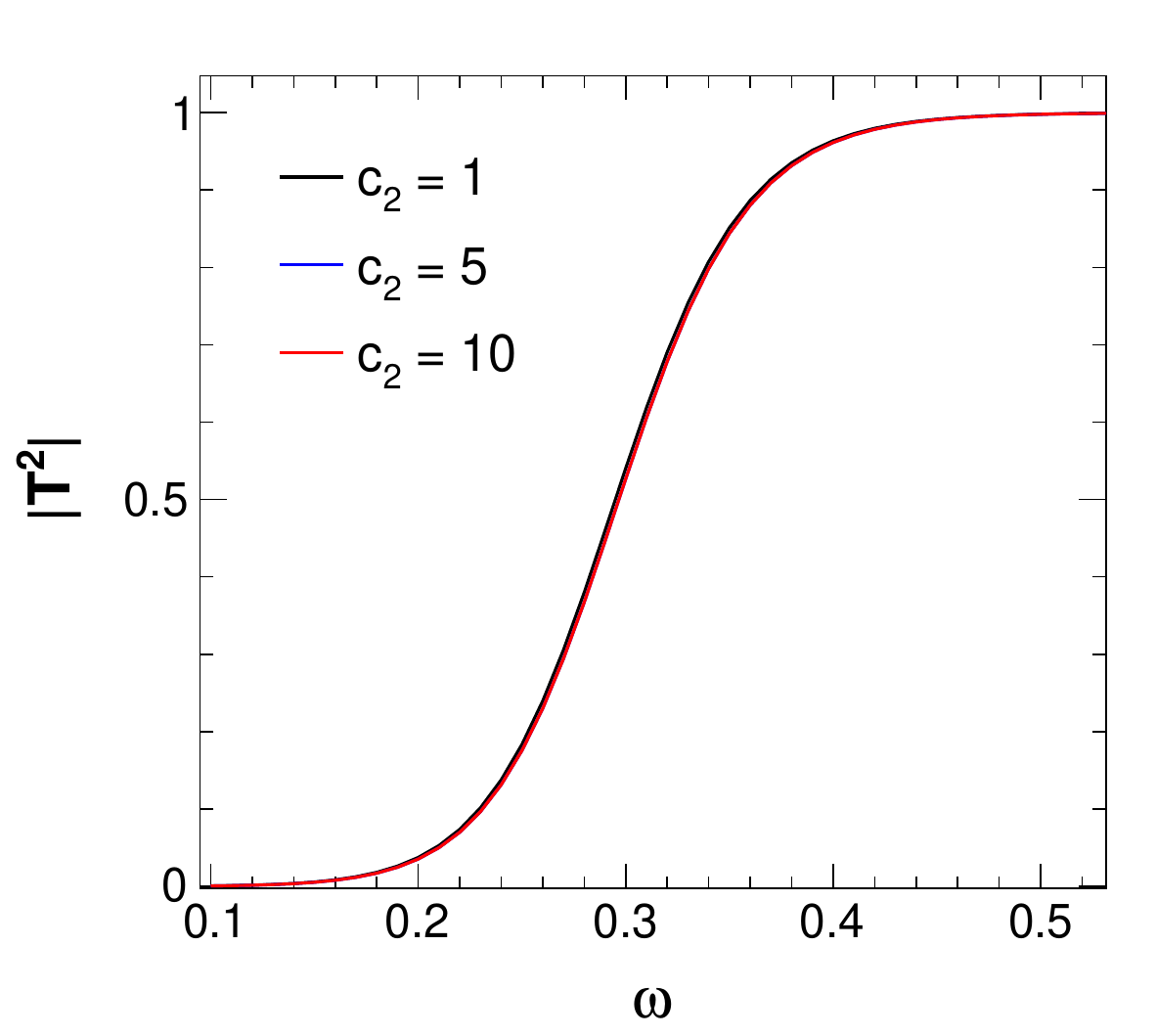}
\caption{Greybody factors versus frequency $\omega$ for different values of 
$m$ with $l=1$ (left plot) and $l=2$ (right plot). We have used $M=n=1$ and 
$c_{2}=1$ in the first two plots and $m=0.1$ for the third plot. The third 
plot shows that greybody factor is independent of parameter $c_2$.}
\label{fig06}
\end{figure}

The greybody factor or the transmission coefficient is a measure of the 
probability that a particle created by quantum processes near the event 
horizon of a black hole will escape to infinity or get absorbed inside the 
black hole. Greybody factor ($T^2$) equal to $1$ means that all the particles 
that are created are able to escape the black hole while lower values of it 
mean that some of them end up inside it. If $T^2=0$, it means that the black 
hole is completely dark and absorbs every single particle. The greybody factor 
has been extensively studied in the literature in various scenarios. We can 
express the reflection and transmission of the particles hitting 
the black hole barrier potential in the following form \cite{77}:
\begin{align}
\psi(x) & = T(\omega) \exp (-i \omega x), \;\; x \rightarrow -\infty,
\label{eqc1}\\[8pt]
\psi(x) & =  \exp (-i \omega x) + R(\omega) \exp(i \omega x),\;\; x \rightarrow +\infty,
\label{eqc2}
\end{align}
where $R(\omega)$ and $T(\omega)$ are respectively reflection and 
transmission coefficients and are functions of frequency $\omega$. WKB 
approximation formula is used to get to the computational form of these 
two coefficients which are presented below \cite{77}:
\begin{align}
|R(\omega)|^2 & =\frac{1}{1+\exp(-2\pi i \tau)},
\label{eqc3}\\[8pt]
|T(\omega)|^2 & = \frac{1}{1+\exp(2\pi i \tau)},
\label{eqc4}
\end{align}
where the parameter $\tau$ is defined in the WKB method as the 
following \cite{77}:
\begin{equation}
\tau=\frac{i (\omega^2 -V_0)}{\sqrt{-2V_0^{''}}}-\Lambda_j.
\label{eqc5}
\end{equation}
Here double primes represent the double derivative of the maximum of the 
effective potential $V_0$ with respect to $x$ and $\Lambda_j$ can be obtained 
from the WKB formula found in Ref.~\cite{75}. In Fig.~\ref{fig06}, we plot the 
greybody factors with respect to frequency $\omega$ for three values of the 
model parameters $m$ considering $n=1$, $c_2=1$ and $M=1$ with multipole $l=1$
(left plot) and $l=2$ (right plot). It is seen that for higher $m$ values
the grebody factor increases faster with respect to $\omega$ than that for 
smaller $m$ values. Also, for a smaller $l$ value ($l=1$), the greybody factor 
increase is more rapid and begins from a smaller $\omega$ value as compared to 
a higher $l$ value ($l=2$). It needs to be mentioned that the values of the 
parameter $c_2$ are found to be insensitive in the variation of the grebody 
factors with respect to $\omega$ as shown in Fig.~\ref{fig06}. 

\section{Shadow of the black hole}  \label{sec5}
Black hole shadow has been extensively studied in the literature as it 
provides good scope to test various theories of gravity and black hole 
physics in extreme gravity regimes. The recent observational data of black 
hole shadow radius has provided the scientific community an opportunity 
to constrain model parameters using these data. In this section, we 
compute the photon sphere and the shadow radius expression and plot the same 
for analysing its dependence on various model parameters. We also try to 
constrain the parameter space with observational data of the EHT group. 

\begin{figure}[h!]
\includegraphics[scale=0.4]{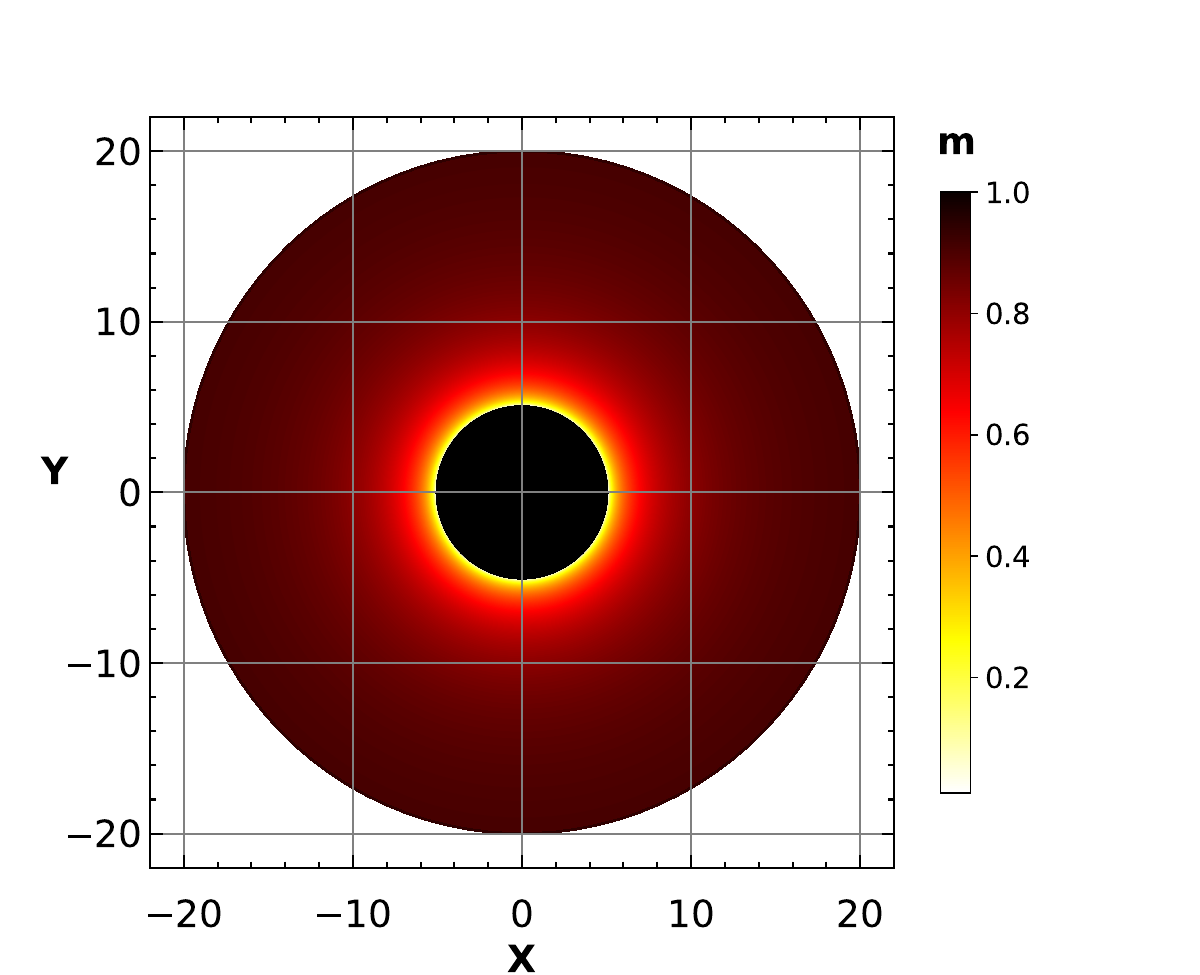}\hspace{-0.5cm}
\includegraphics[scale=0.4]{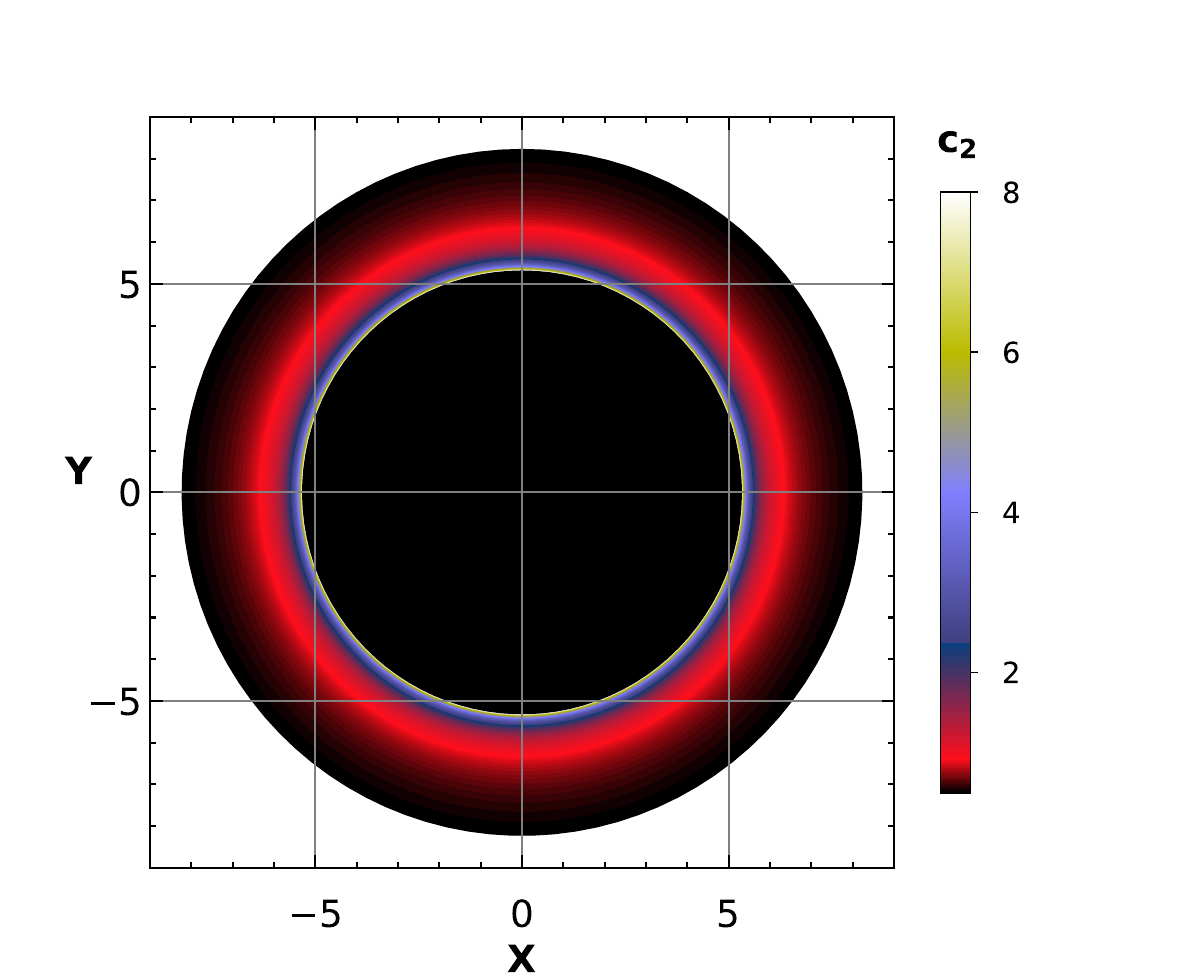}
\caption{Stereoscopic projection of shadow radius in terms of celestial 
coordinates. The left plot is for the variation of $m$ with parameters 
$c_{2}=M=n=1$ and the right plot is for the variation of $c_2$ with 
$M=n=1$ and $m=0.1$}.
\label{fig07}
\end{figure}

The simple condition to determine the photon sphere radius of a black hole 
in spherical symmetry consideration is given by the following relation 
\cite{71-1,71-2}:
\begin{equation}
2-\frac{r N'(r)}{N(r)}=0.
\label{eqc1}
\end{equation}
Using the form of $N(r)$ from Eq.~\eqref{eqa20}, we solve the Eq.~\eqref{eqc1} 
for $r$ to get the photon sphere radius as
\begin{equation}
r_{ph}=3M.
\label{eqc2}
\end{equation}
From the photon radius, we can derive the shadow radius as follows:
\begin{equation}
r_{sh}=\frac{r_{ph}}{\sqrt{N(r)}|_{r\rightarrow r_{ph}}}=\frac{3 M}{\sqrt{\frac{3}{4} m^2 M^2 \left(\frac{n-2}{\text{2\,c2}}\right)^{1/n}+\frac{1}{3}}}.
\label{eqc3}
\end{equation}
Obviously, the shadow radius depends on all three 
model parameters associated with $N(r)$. It is also evident that when model 
parameters $m=n=c_2=0$, we recover the standard $r_{sh}=3\sqrt{3}M$ which is 
the shadow radius for the Schwarzschild black hole. Now, for the 2-D 
stereoscopic projection of shadow radius, we define celestial coordinates $X$ 
and $Y$ as given by \cite{71-1,71-2}
\begin{align}
 X & =\lim_{r_{0}\rightarrow\infty}\left(-\,r_{0}^{2}\sin\theta_{0}\left.\frac{d\phi}{dr}\right|_{r_{0}}\right),\\[5pt]
Y & =\lim_{r_{0}\rightarrow\infty}\left(r_{0}^{2}\left.\frac{d\theta}{dr}\right|_{(r_{0},\theta_{0})}\right).
\label{eqc4}
\end{align}
Here $\theta_0$ is the observer's angular position with regards to the plane 
of the black hole. In Fig.\ref{fig07}, we show the variation of the shadow 
radius with parameters $m$ and $c_2$. It is seen from the left plot that with 
an increase in $m$, the shadow radius increases. In the right plot, it is 
evident that the shadow radius decreases with increasing $c_2$ 
values. Thus, the parameters $m$ and $c_2$ have opposite influences on the 
shadow radius. 

In order to constrain the parameters of the model, we shall employ the 
technique mentioned in Ref.~\cite{71-1}. We briefly present some 
important steps in this direction. The main point of the methodology is that 
we compare the observed angular radius of the Sgr A* black hole as captured 
by the EHT group recently with the theoretically calculated shadow radius 
from the expression \eqref{eqc3} by constraining the model parameters. This 
requires the prior value of the mas-to-distance ratio for Sgr A*. Another 
feature that is required for this method is the calibration factor that 
correlates the observed to the calculated shadow radius. This method has been 
used to constrain model parameters in the literature \cite{78,79,80} and we 
shall follow the same route. 

\begin{figure}[h!]
\includegraphics[scale=0.45]{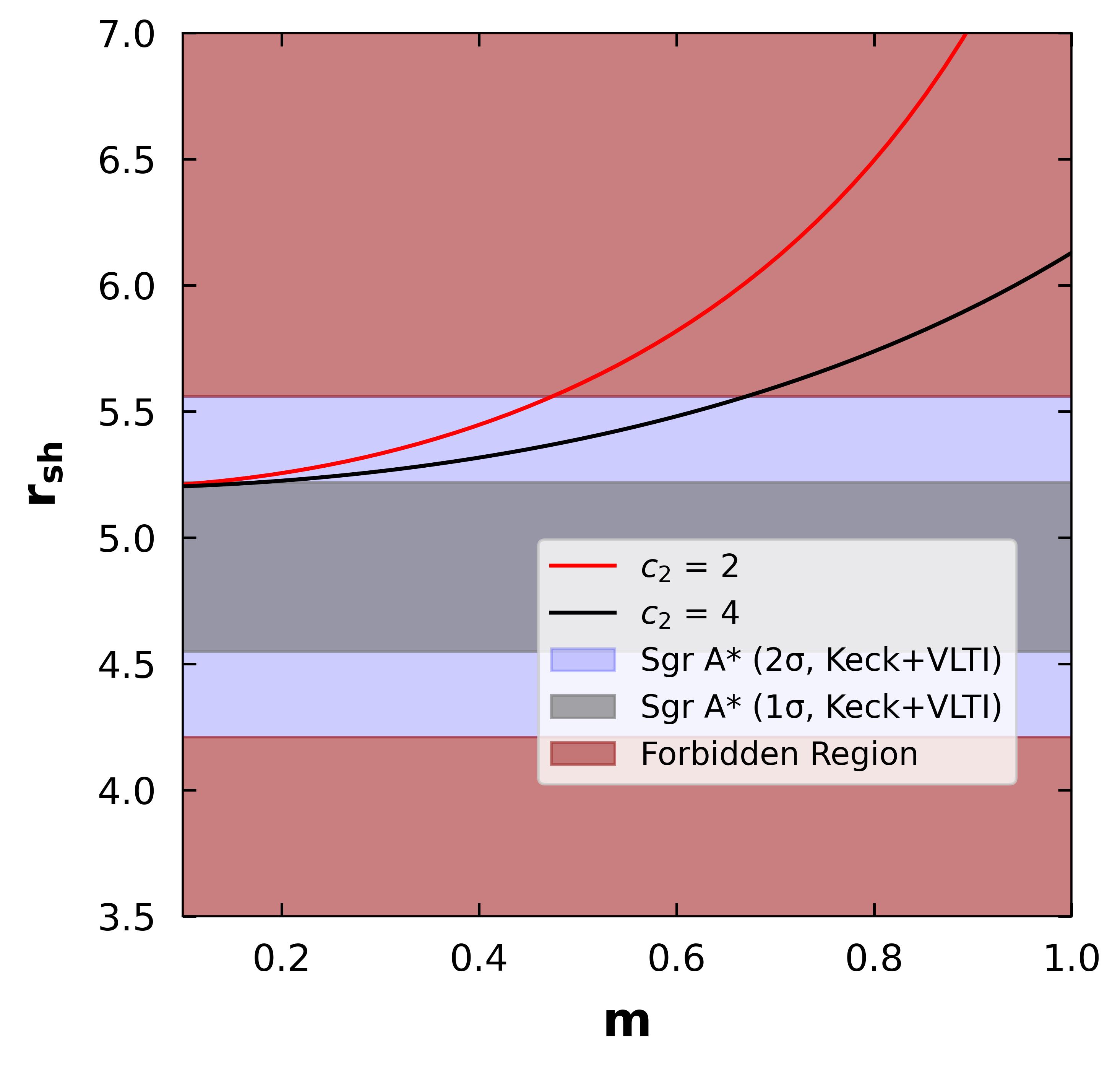}\hspace{0.5cm}
\includegraphics[scale=0.45]{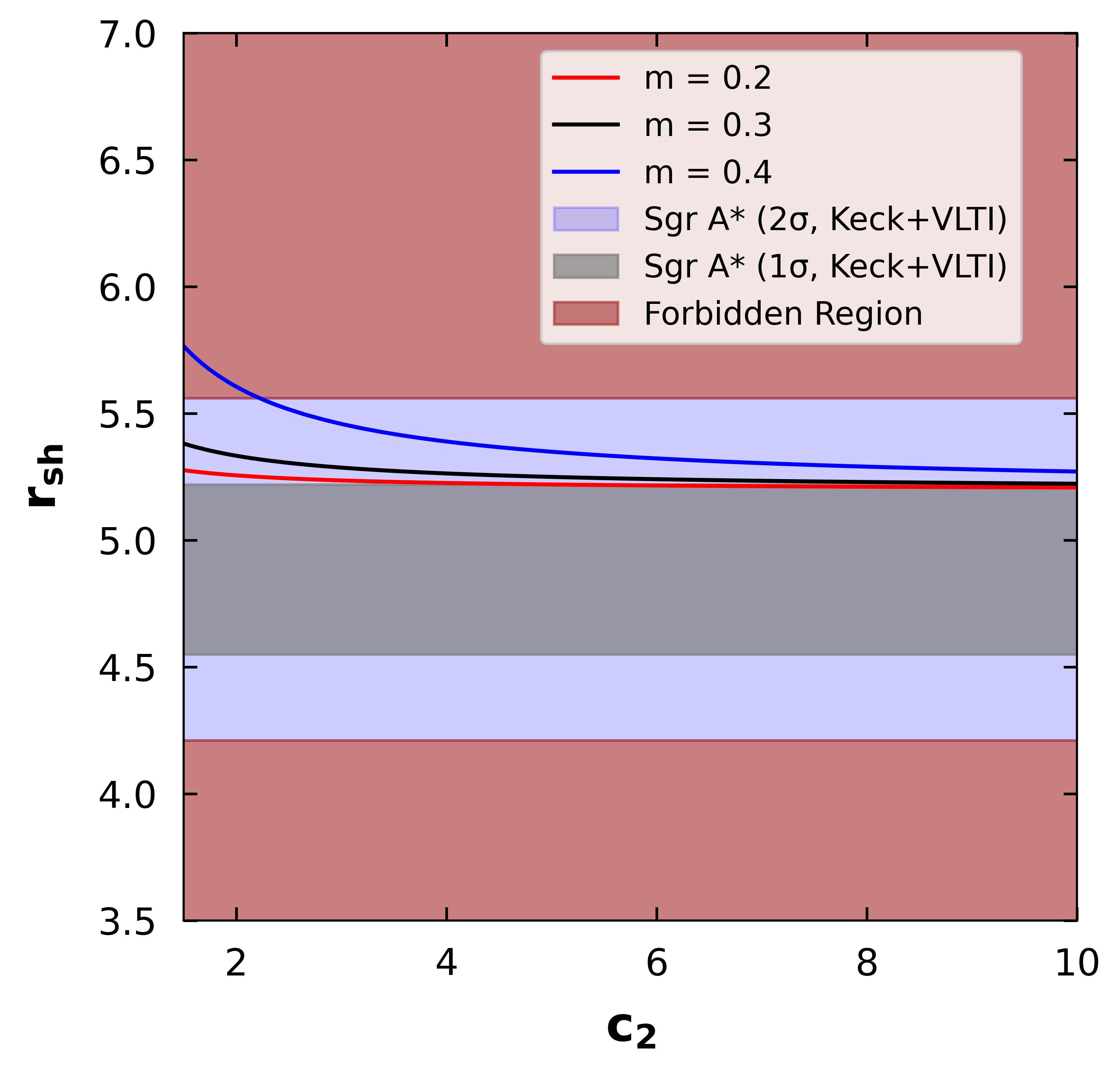}
\vspace{-0.2cm}
\caption{Shadow radius versus parameter $m$ and $c_{2}$ have been plotted in 
the background of Keck and VLTI constrains \cite{71-1} from observations of 
Sgr A*. We have chosen $M=1$ and $n=1$ for these plots. The red portion 
represents the zone forbidden by Keck-VLTI observation.}
\label{fig08}
\end{figure}

A new parameter $\delta$ defined by the EHT group to refer to the fractional 
deviation between the observed shadow radius $r_{s}$ and shadow radius of a 
Schwarzschild black hole $r_{sch}$ is \cite{71-1}
\begin{equation}
\delta=\frac{r_{s}}{r_{sch}}-1=\frac{r_{s}}{3\sqrt{3}M}-1.
\label{eqc5}
\end{equation}
This parameter was estimated by the Keck and VLTI measurements as \cite{71-1}
\begin{align*}
\textrm{Keck}: \delta=- \,0.04^{+0.09}_{-0.10}\\[5pt] 
\textrm{VLTI}: \delta=-\, 0.08^{+0.09}_{-0.09}
\end{align*}
For simplification, we shall adopt the mean of the two observations as 
considered in Ref.~\cite{71-1} in the rest of the work, which is 
\begin{equation}
\delta=-\,0.060 \pm 0.065.
\label{eqc6}
\end{equation}
This leads to $\delta$ parameter's 1$\sigma$ and $2\sigma$ intervals as
\begin{align}
-\,0.125 \lesssim & \,\delta \lesssim 0.005\, (1\sigma),
\label{eqc7}\\[5pt]
%\end{equation}
%\begin{equation}
-\,0.190 \lesssim & \,\delta\lesssim 0.070\,(2\sigma).
\label{eqc8}
\end{align}
It is found that the bounds \eqref{eqc7} and \eqref{eqc8} when imposed 
upon Eq.~\eqref{eqc5} give the bounds on $r_{sh}$ as follows \cite{71-1}:
\begin{align}
4.55 \lesssim &\, r_{sh}/M \lesssim 5.22\, (1\sigma),
\label{eqc9}\\[5pt]
%\end{equation}
%\begin{equation}
4.21 \lesssim &\, r_{sh}/M \lesssim 5.56\, (2\sigma).
\label{eqc10}
\end{align}
We plot the shadow radius with the bounds imposed by the observations of Keck 
and VLTI in Fig.~\ref{fig08}. The left plot shows that the shadow radius 
increases with increasing $m$ values as found in Fig.~\ref{fig07}. It shows 
that for smaller values of $c_{2}$, the shadow radius quickly moves to the 
forbidden region. With increasing $c_{2}$ values, the shadow radius within 
the allowed region increases. In the right plot, the shadow radius is plotted 
versus $c_{2}$ which shows that the shadow radius decreases with increasing 
$c_{2}$ values as observed earlier. The plots are within the 2$\sigma$ allowed 
region in this case with the exception of larger $m$ and smaller $c_{2}$ 
values as clearly visible from the plot. It is quite evident that the 
constraints imposed are not rigid but depend on the range of values of the 
parameters of the model.

This method of constraining parameters of a theory has been adopted in the 
literature \cite{78,79} and by the EHT group themselves \cite{80} and provides 
a robust way of constraining parameters. But in cases of model parameters 
exceeding one, we need some supporting constraining methods so that one 
parameter can be cornered and rigorous constraints can be obtained. However, 
we leave this as a future extension of the work.

\section{Conclusion} \label{sec6}

In this work, we derive novel black hole solutions in the framework of 
Hu-Sawicki gravity. We plot the metric function versus $r$ for various values 
of model parameters and encountered two horizons of the black hole. It is seen 
that higher $m$ values cause the horizon radius to shrink while opposite trend 
is observed for parameter $c_2$. We then analyse the QNMs of the novel black 
hole solution with 6th order WKB approximation. The amplitude increases with 
increase in $c_2$ while it decreases with $m$. The damping decreases with 
increasing $m$ while it increases slightly with $c_2$. This trend can be 
realised from the tabulated QNM data in Table \ref{tab01}. It is evident 
that the QNM frequencies are affected by the model parameters. The associated 
error is found to be around $10^{-4}$ to $10^{-5}$ in some cases.  

Thermodynamic temperature associated with the black hole is investigated 
and it is found to decrease with $r_H$ in all cases. The temperature can also 
become negative, suggesting the possibility of formation of ultra-cold black 
hole. The heat capacity of the black hole has been computed and plotted 
the same with respect to the horizon radius. It is seen that the heat capacity 
takes negative values for most of the parameter space, indicating 
thermodynamic instability of the obtained black hole solution.
The greybody factors are also computed, specially the transmission coefficients 
with respect to frequency $\omega$ and the dependence of the model parameter $m$ 
is studied. Higher $m$ results in swifter increase in the greybody factors towards 
saturation value of 1. It is noteworthy that increasing the multipole $l$ lowers 
the rate of increase of greybody factors and saturation is achieved at higher $\omega$. 

The photon radius and the shadow radius asociated with the spherically 
symmetric black hole spacetime are then studied. We presented the stereographic 
projection of the shadow in celestrial coordinate system and using contour-type 
feature, showed the variation of the shadow radius with increasing model 
parameters $m$ and $c_{2}$. Using the already established constraints on 
$r_{sh}$ by Keck and VLTI observations, we constrain our model parameters 
using a well proven scheme. The parameter $m$ is roughly constrained to be 
less that $\sim$0.5 while parameter $c_{2}$ is constrained to be greater than 
$\sim$2, as can be seen from Figure \ref{fig08}.

The recent technical advancements made in the fields of astrophysics and observational 
astronomy has made the present era very suitable for theoretical physicists to constrain 
and test fundamental theories and models, which was not possible untill a decade back. 
With ground-breaking leaps in the form of LIGO-Virgo team's observation of Gravitational 
Waves in 2015 along with the first-ever image of the black hole M87* and later that of 
Sgr A*, scientists plan to 
further enhance sensitivity of the present detectors as well as new ambitious projects 
like the space-based LISA project and the Einstein Telescope are already in the 
planning stages. As a future scope of this work, we can analyse other viable models of 
gravity like $f(R,T)$ and $f(Q)$ and work on new black hole solutions as well as rotating 
Kerr-type solutions can also be explored. The study of black hole shadows surely holds 
a lot of potential in constraining fundamental physics and it certainly deserves 
further investigation.

%%%%%%%%%%%%%%%%%%%%%%%%%%%%%%%%%%%%%%%%%%%%%%%%%%%
\section*{Acknowledgements}  UDG is thankful to the Inter-University Centre for
Astronomy and Astrophysics (IUCAA), Pune, India for awarding the Visiting 
Associateship of the institute.
%%%%%%%%%%%%%%%%%%%%%%%%%%%%%%%%%%%%%%%%%%%%%%%%%%%%%

\end{document}